\documentclass[journal,letterpaper,twoside]{IEEEtran}

\usepackage{amsmath}
\usepackage{amssymb}
\usepackage{amsfonts}
\usepackage{amscd}
\usepackage{amsthm}
\usepackage[noadjust]{cite}
\usepackage{textcomp}
\usepackage{multirow}
\usepackage{amsbsy}
\usepackage{graphicx}
\usepackage{hhline}
\usepackage{xfrac}

\newtheorem{lemma}{Lemma}

\theoremstyle{definition}
\newtheorem{example}{Example}
\newtheorem{definition}{Definition}

\makeatletter
\renewcommand*\env@matrix[1][c]{\hskip -\arraycolsep
  \let\@ifnextchar\new@ifnextchar
  \array{*\c@MaxMatrixCols #1}}
\makeatother

\newcommand{\La}{\Lambda}
\newcommand{\Rb}{\mathbb{R}}
\newcommand{\Zb}{\mathbb{Z}}
\newcommand{\Xc}{\mathcal{X}}
\newcommand{\Wc}{\mathcal{W}}

\newcommand{\Sc}{\bar{S}} 

\renewcommand{\mod}{{\rm~mod~}}

\title{Index Codes for the Gaussian Broadcast Channel \\ using Quadrature Amplitude Modulation}
\author{
Lakshmi~Natarajan, Yi~Hong, and Emanuele~Viterbo%
\thanks{The authors are with the Department of Electrical and Computer Systems Engineering, Monash University, Clayton, VIC 3800, Australia (email: \{lakshmi.natarajan, yi.hong, emanuele.viterbo\}@monash.edu).}%
\thanks{\copyright~2015 IEEE. Personal use of this material is permitted. Permission from IEEE must
be obtained for all other uses, including reprinting/republishing this material for
advertising or promotional purposes, collecting new collected works for resale or
redistribution to servers or lists, or reuse of any copyrighted component of this work
in other works.}
}

\begin{document}

\maketitle

\begin{abstract}
\boldmath
We propose index codes, based on multidimensional QAM constellations, for the Gaussian broadcast channel, where every receiver demands all the messages from the source. The efficiency with which an index code exploits receiver side information in this broadcast channel is characterised by a code design metric called \emph{side information gain}. The known index codes for this broadcast channel enjoy large side information gains, but do not encode all the source messages at the same rate, and do not admit message sizes that are powers of two.
The index codes proposed in this letter, which are based on linear codes over integer rings, overcome both these drawbacks and yet provide large values of side information gain.
With the aid of a computer search, we obtain QAM index codes for encoding up to $5$ messages with message sizes $2^m$, \mbox{$m \leq 6$}.
We also present the simulated performance of a new $16$-QAM index code, concatenated with an off-the-shelf LDPC code, which is observed to operate within $4.3$~dB of the broadcast channel capacity.
\end{abstract}


\section{Introduction} \label{sec:1}

\IEEEPARstart{C}{oding} for broadcast channels, where receivers know some part of the transmitted messages a priori, is called \emph{index coding} and is well-known for noiseless binary broadcast channels~\cite{YBJK_IEEE_IT_11,ALSWH_FOCS_08,RSG_IEEE_IT_10}.
In the case of noisy binary broadcast, the index codes of~\cite{DSC_IT_13} provide equal error correcting capability at all receivers and exploit the receiver side information to enhance the code rate, while the codes of~\cite{XFKC_CISS_06,BaC_ITW_11,MLV_PIMRC_12} transform side information into improvements in error performance.
The capacity of general index coding over Gaussian broadcast channel is unknown, but information theoretic results are available for some special cases~\cite{KrS_ITW_07,Wu_ISIT_07,SiC_ISIT_14,AOJ_ISIT_14,Tun_IEEE_IT_06}.
Separation-based coding schemes using a (noiseless) index code and a broadcast channel code are, in general, sub-optimal, since the channel decoders do not utilize the receiver side information, and the channel coding rate is limited by the receiver with the worst signal-to-noise ratio. This motivates schemes that perform index coding at the physical layer.

Lattice based codes were proposed in~\cite{NHV_arxiv_14} for the special case of index coding over the Gaussian broadcast channel where the transmitter has $K$ independent messages, each receiver knows some subset of the $K$ messages a priori, and every receiver demands all the messages at the source. 
These index codes are designed to convert receiver side information into apparent ${\sf SNR}$ gains. The minimum distance of the effective code perceived by a receiver is a function of the index subset \mbox{$S \subset \{1,\dots,K\}$} of the messages available at the receiver as side information.
The \emph{side information gain} of a code is a metric that measures the efficiency with which receiver side information is converted to actual coding gain~\cite{NHV_arxiv_14}. The index codes of~\cite{NHV_arxiv_14} provide large side information gains, and they can be concatenated with outer channel codes to improve coding gain against channel noise.
These index codes, however, suffer from two practical drawbacks: \emph{(i)}~they do not encode all messages at equal rate, and \emph{(ii)}~they do not admit message sizes that are powers of $2$.


In this letter, we present the first class of index codes for this special case of Gaussian broadcast channel that encode all the messages with equal rate (Section~\ref{sec:3}). These new index codes allow messages of arbitrary sizes, including sizes that are powers of $2$. The proposed index codes are multidimensional QAM constellations whose points are labelled with message symbols using the framework of linear codes over the ring $\Zb_M$ of integers modulo $M$. 
Using a computer search, we obtain QAM index codes with large side information gains for message sizes $2^m$, \mbox{$m \leq 6$}, and number of messages \mbox{$K \leq 5$}.
We also present simulation results on the performance of a QAM index code when used as a modulation scheme in a system employing an outer channel code (Section~\ref{sec:4}). We observe that the new $16$-QAM index modulation scheme for \mbox{$K=2$} messages, when encoded with an off-the-shelf rate-$\sfrac{1}{2}$ LDPC code, performs $4.3$~dB away from capacity in the Gaussian broadcast channel at $10^{-4}$ bit error rate.


\section{Index codes for Gaussian Broadcast Channel} \label{sec:2}

We consider a non-fading Gaussian broadcast channel with single-antenna terminals, where every receiver demands $K$ independent messages from the transmitter, denoted by $w_1,\dots,w_K$ that assume values from $\Wc_1,\dots,\Wc_K$, respectively. The transmitter operates under an average power constraint, the receivers experience additive white Gaussian noise (with possibly different noise powers), and each receiver has prior knowledge of some subset of the $K$ messages as side information. 
An $n$--dimensional \emph{index code} $(\rho,\Xc)$ for this Gaussian broadcast channel consists of a channel code \mbox{$\Xc \subset \Rb^n$} and an encoding function \mbox{$\rho: \Wc_1 \times \cdots \times \Wc_K \to \Xc$}. The rate of transmission of the $k^{\text{th}}$ message is $R_k=\sfrac{1}{n} \log_2 |\Wc_k|$ bits per dimension (b/dim). 
A receiver that has the prior knowledge of the symbols \mbox{$\pmb{w}_S=(w_k,k \in S)$}, \mbox{$S \subsetneq \{1,\dots,K\}$}, and experiences a signal-to-noise ratio of ${\sf SNR}$ is denoted by $({\sf SNR},S)$. 
We are interested in codes that provide good error performance (versus ${\sf SNR}$) for every \mbox{$S \subsetneq \{1,\dots,K\}$}, or equivalently, for \mbox{$2^K-1$} receivers, one corresponding to each $S \subsetneq \{1,\dots,K\}$.

Consider the channel output \mbox{$\pmb{y} = \rho(w_1,\dots,w_K) + \pmb{z}$} at a generic receiver $({\sf SNR},S)$, where \mbox{$\pmb{z} \in \Rb^n$} is the additive Gaussian noise with variance $\sfrac{1}{\sf SNR}$ per dimension. 
A receiver with no side information, i.e. with \mbox{$S=\varnothing$}, decodes $\pmb{y}$ to \mbox{$\arg \min_{\pmb{x} \in \Xc} \|\pmb{y}-\pmb{x}\|$}. The minimum Euclidean distance \mbox{$d_0=\min \{ \|\pmb{x}_1 - \pmb{x}_2\|\, \vert  \,\pmb{x}_1,\pmb{x}_2 \in \Xc, \pmb{x}_1 \neq \pmb{x}_2 \}$} between any pair of points in $\Xc$ determines the error performance at this receiver. A receiver with \mbox{$S \neq \varnothing$} has prior knowledge of the value of the message vector $\pmb{w}_S$. Given the information \mbox{$w_k=a_k$}, \mbox{$k \in S$}, written concisely as \mbox{$\pmb{w}_S=\pmb{a}_S$}, this receiver generates a subcode \mbox{$\Xc_{\pmb{a}_S} \subset \Xc$} by expurgating all codewords in $\Xc$ with \mbox{$\pmb{w}_S \neq \pmb{a}_S$}, and decodes $\pmb{y}$ to the closest point in $\Xc_{\pmb{a}_S}$. 
Let $d_{\pmb{a}_S}=\{\|\pmb{x}_1-\pmb{x}_2\|\,\vert\,\pmb{x}_1,\pmb{x}_2 \in \Xc_{\pmb{a}_S}, \pmb{x}_1\neq\pmb{x}_2\}$ be the minimum Euclidean distance of $\Xc_{\pmb{a}_S}$, and $d_S=\min_{\pmb{a}_S} d_{\pmb{a}_S}$.
The average error performance and coding gain at this receiver are determined by $d_S$.
The asymptotic additional ${\sf SNR}$ gain due to the knowledge of $\pmb{w}_S$ is thus \mbox{$10 \log_{10} \left( \sfrac{d_S^2}{d_0^2} \right)$~dB}. This squared distance gain must be measured against the amount of side information in $\pmb{w}_S$, or equivalently, against the \emph{side information rate} \mbox{$R_S \triangleq \sum_{k \in S}R_k$~b/dim}.
The \emph{side information gain}~\cite{NHV_arxiv_14} of the code $(\rho,\Xc)$, defined as
\begin{equation} \label{eq:Gamma}
\Gamma \triangleq \min_{\varnothing \subsetneq S \subsetneq \{1,\dots,K\}} \frac{10 \log_{10} \left( \sfrac{d_S^2}{d_0^2} \right)}{R_S} \text{ dB/b/dim},
\end{equation}
is the minimum additional coding gain available from each bit per dimension of side information for any $S$.
The prior knowledge of $\pmb{w}_S$ provides an asymptotic ${\sf SNR}$ gain of at least \mbox{$\Gamma \times R_S$~dB} over the performance of $\Xc$ with no side information. Hence, $(\rho,\Xc)$ is a good index code if \emph{(i)}~$\Xc$ is a good channel code for the traditional single user AWGN channel, i.e., for a receiver with \mbox{$S=\varnothing$}, and \emph{(ii)}~$\Gamma$ is large, so as to maximize the minimum gain from side information for receivers with \mbox{$S \neq \varnothing$}.


To motivate our work, we now show an example of a new index code using $16$-QAM, that encodes two $4$-ary message symbols with equal rate, and provides \mbox{$\Gamma \approx 6$~dB/b/dim}.

\begin{example} \label{ex:16QAM_1}
\begin{figure}[!t]
\centering
\includegraphics[totalheight=2.0in,width=2.0in]{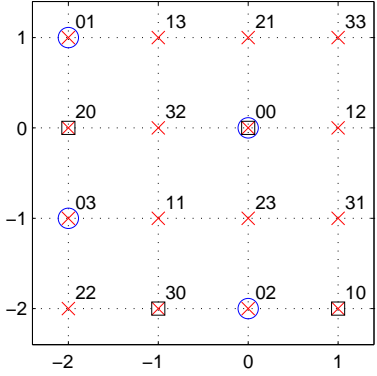}
\caption{The labelling scheme for the $16$-QAM index code. The four points forming the subcode corresponding to the side information \mbox{$w_1=0$} are highlighted with circles. The subcode for \mbox{$w_2=0$} is marked with squares.}
\label{fig:16QAM_labels}
\vspace{-3mm}
\end{figure} 
Let \mbox{$K=2$}, and number of receivers be \mbox{$2^K-1=3$}, with the corresponding side information index sets \mbox{$S=\varnothing,\{1\},\{2\}$},  respectively. Let {$\Wc_1=\Wc_2=\{0,1,2,3\}$}, \mbox{$n=2$} and $\Xc$ be the $16$-QAM constellation, then \mbox{$R_1=R_2=1$~b/dim}. Fig.~\ref{fig:16QAM_labels} depicts the new code, where each of the $16$ points $\pmb{x}$ is labelled with the corresponding message tuple \mbox{$\rho^{-1}(\pmb{x})=(w_1,w_2)$}.
The receiver with \mbox{$S=\varnothing$} must decode both $w_1,w_2$, and hence, it decodes the received vector to nearest point in $\Xc$. The error performance at this receiver is that of the $16$-QAM signal set. Let \mbox{$w_1=0$}, then the receiver with \mbox{$S=\{1\}$} knows that the transmit vector is one of the four points corresponding to \mbox{$w_1=0$} (marked with circles in Fig.~\ref{fig:16QAM_labels}), and hence, its decoder restricts its choice of candidate codewords to these four points.
Observe that the minimum Euclidean distance between these four points is twice the minimum Euclidean distance $d_0$ of $\Xc$. The minimum distance corresponding to each of the other three values of $w_1$ is also $2d_0$, and hence, \mbox{$d_S=2d_0$} for \mbox{$S=\{1\}$}. It is easy to check that \mbox{$d_S=2d_0$} for \mbox{$S=\{2\}$} as well. Thus, the error performance at the two receivers, corresponding to \mbox{$S=\{1\},\{2\}$}, respectively, is approximately \mbox{$10\log_{10}(2^2)\approx 6$~dB} better than that of the receiver with \mbox{$S=\varnothing$}. 
Since \mbox{$R_S=1$~b/dim} for $S=\{1\},\{2\}$, from~\eqref{eq:Gamma}, the side information gain of this code is $10\log_{10}(2^2) \approx 6$~dB/b/dim. \hfill\IEEEQED
\end{example}

\section{QAM constellations for index coding} \label{sec:3}

In this section, we present multidimensional QAM constellations for index coding using linear codes over the ring of integers modulo $M$. 
For even and odd values of $M$, let $\Zb_M$ denote the sets 
${\textstyle \left\{-\frac{M}{2},-\frac{M-2}{2},\dots,0,\dots,\frac{M-2}{2}\right\}}$ and ${\textstyle \left\{-\frac{M-1}{2},-\frac{M-3}{2},\dots,0,\dots,\frac{M-1}{2}\right\}}$,
respectively. For any \mbox{$a \in \Zb$}, let \mbox{$a \mod M$} be the unique remainder of $a$ in $\Zb_M$ when divided by $M$. With addition and multiplication performed modulo $M$, the set $\Zb_M$ has the structure of a commutative ring. The $\mod M$ operation satisfies the property that for any \mbox{$x \in \Zb$}, \mbox{$|x \mod M| \leq |x|$}. The set $\Zb_M^n$ of all $n$-tuples is a module over $\Zb_M$ with addition and scalar multiplication performed component-wise. 
Similar to the scalar case, we have \mbox{$\|\pmb{x} \mod M \| \leq \|\pmb{x}\|$} for every \mbox{$\pmb{x} \in \Zb^n$}.

A \emph{unit} is an element of a ring with a multiplicative inverse, and the set of all units of a ring form a multiplicative group. In the case of $\Zb_M$, the units are precisely the elements that are relatively prime with $M$ in $\Zb$, i.e.,
\mbox{$U(\Zb_M) = \left\{ a \in \Zb_M \, \vert \, \gcd(a,M) = 1 \text{ in } \Zb \right\}$},
where $\gcd$ denotes the greatest common divisor. When $M$ is a power of $2$, $U(\Zb_M)$ is the set of all odd integers in $\Zb_M$.

Assuming \mbox{$|\Wc_1|=\cdots=|\Wc_K|=M$}, we identify each alphabet $\Wc_k$ with the ring $\Zb_M$. We consider $\Zb_M$--linear encoding of the $K$ messages where the code length equals the number of messages, i.e., \mbox{$n=K$}, and the subcode associated with each message is of rank $1$. The $k^{\text{th}}$ subcode \mbox{$\Xc_k = \left\{ w_k\pmb{c}_k \mod M \, | \, w_k \in \Zb_M \right\}$}, corresponding to the message $w_k$, is generated by a single vector \mbox{$\pmb{c}_k \in \Zb_M^K$}. 

\begin{definition} \label{def:linear_index_code} 
A \emph{$\Zb_M$-linear index code} for $K$ messages consists of a set of $K$ generators \mbox{$\pmb{c}_1,\dots,\pmb{c}_K \in \Zb_M^K$}, such that the linear encoder 
\mbox{$\pmb{x} = \rho(w_1,\dots,w_K) = \sum_{k=1}^{K} w_k \pmb{c}_k \mod M$}
is injective.
\end{definition}

The injectivity of $\rho$ in Definition~\ref{def:linear_index_code} ensures unique decodability of messages at a receiver with no side information. Since the message space \mbox{$\Wc_1 \times \cdots \times \Wc_K=\Zb_M^K$}, injectivity of $\rho$ implies that \mbox{$\Xc=\Zb_M^K$}. In order to transmit the signal, we embed the codeword \mbox{$\pmb{x} \in \Zb_M^K$} into the Euclidean space $\Rb^K$ using the natural map. Hence, the minimum distance with no side information is \mbox{$d_0=1$}. 
The linear index code can be viewed as a labelling of the multidimensional QAM constellation $\Zb_M^K$, where each constellation point $\pmb{x}$ is associated with the message tuple $(w_1,\dots,w_K)=\rho^{-1}\left(\pmb{x}\right)$. Note that $\pmb{x}$ may be translated by a fixed offset prior to transmission to minimize the transmit power.

A linear index code is fully characterized by the matrix \mbox{$\pmb{C} \in \Zb_M^{K \times K}$} whose rows are the $K$ generators $\pmb{c}_1,\dots,\pmb{c}_K$. The encoding matrix $\pmb{C}$ defines a linear transformation from the message space \mbox{$\Zb_M^K$} to the space \mbox{$\Xc=\Zb_M^K$} of codewords. Thus, the encoder map $\rho$ is injective if and only if $\pmb{C}$ is invertible over $\Zb_M$, i.e., \mbox{$\det(\pmb{C}) \in U(\Zb_M)$}. 

\begin{example}[$16$-QAM] \label{ex:16QAM_2}
Consider \mbox{$M=4$}, \mbox{$K=2$} and the two generators \mbox{$\pmb{c}_1= (1,-2)$} and \mbox{$\pmb{c}_2= (-2,1)$}. The encoder is
$\pmb{x} =  w_1\pmb{c}_1 + w_2 \pmb{c}_2 \mod 4 = (w_1 - 2w_2, -2w_1 + w_2) \mod 4$,
and the encoding matrix is
\mbox{$\small \pmb{C} = \begin{pmatrix} \pmb{c}_1 \\ \pmb{c}_2 \end{pmatrix} = \begin{pmatrix}[r] 1 & -2 \\ -2 & 1 \end{pmatrix}$}.
Since \mbox{$\det(\pmb{C})=-3 \mod 4 =1$} is a unit in $\Zb_4$, this code is uniquely decodable. The resulting index code is the $16$-QAM labelling scheme illustrated in Example~\ref{ex:16QAM_1} and Fig.~\ref{fig:16QAM_labels}. \hfill\IEEEQED
\end{example}

\subsection{Side information gain}

All the $K$ messages have the same transmission rate \mbox{$R_k = \sfrac{1}{K} \log_2 M$~b/dim}. The side information rate at the receiver $({\sf SNR},S)$ is \mbox{$R_S = \sum_{k \in S} R_k = \frac{|S|}{K} \log_2 M$~b/dim.}
We now relate the minimum distance $d_S$ to the length of the shortest vector of a certain lattice. 
This allows us to numerically compute the value of $d_S$, and hence $\Gamma(\Xc)$, using efficient algorithms available for calculating the shortest vectors in lattices~\cite{FiP_AMS_85}. 
Let $\Sc$ denote the complement of the set $S$.
For any \mbox{$S \subset \{1,\dots,K\}$}, the subcode generated by $w_k$, \mbox{$k \in \Sc$}, is
\mbox{$\Xc_{\Sc} = \left\{ \sum_{k \in \Sc} w_k\pmb{c}_k \mod M \, \Big\vert \, w_k \in \Zb_M \right\}$}.
Consider 
\begin{equation*}
\textstyle \La_{\Xc_{\Sc}}=\Xc_{\Sc} + M\Zb^K = \left\{\pmb{x} + M\pmb{u} \, \vert \, \pmb{x} \in \Xc_{\Sc}, \pmb{u} \in \Zb^K\right\},
\end{equation*} 
which is known as the \emph{Construction~A lattice}~\cite{CoS_Springer_99} of the linear code $\Xc_{\Sc}$.
The lattice $\La_{\Xc_{\Sc}}$ is generated by $\pmb{c}_k$, $k \in \Sc$, and the $K$ rows of $M\pmb{I}_K$. A basis for $\La_{\Xc_{\Sc}}$ can be efficiently computed from this set of generators, for example, using an algorithm based on LLL reduction~\cite{BuP_LecNotes_CompSc_87}.
For any set of points in $\Rb^K$, let $d_{\min}(\cdot)$ denote the minimum Euclidean distance between any two distinct points in the set. 
For a lattice $\La$, $d_{\min}(\La)$ equals the length of its shortest vector. 

\begin{lemma} \label{lem:dmin_lattice}
If $\La_{\Xc_{\Sc}}$ contains a shortest vector $\pmb{w}$ such that \mbox{$\pmb{w} \notin M\Zb^K$}, then $d_S=d_{\min}\left( \La_{\Xc_{\Sc}} \right)$; else $d_S \geq M$.
\end{lemma}
\begin{IEEEproof}
Let the side information at the receiver $({\sf SNR},S)$ be \mbox{$\pmb{w}_S=\pmb{a}_S$}. Then the subcode $\Xc_{\pmb{a}_S}$ to be decoded is
\begin{equation*}
\textstyle  \left\{ \sum_{k \in S} a_k \pmb{c}_k + \sum_{k \in \Sc} w_k \pmb{c}_k \mod M  \Big\vert  w_k \in \Zb_M, k \in \Sc \right\},
\end{equation*} 
that equals \mbox{$\pmb{t} + \Xc_{\Sc} \mod M$}, 
where \mbox{$\pmb{t}=\sum_{k \in S} a_k \pmb{c}_k \mod M$} is known at the receiver. Since the modulo operation is equivalent to the addition of an appropriate vector from $M\Zb^K$, we have 
\begin{equation*}
\Xc_{\pmb{a}_S}=\pmb{t}+\Xc_{\Sc} \mod M \subset \pmb{t} + \Xc_{\Sc}+M\Zb^K=\pmb{t}+\La_{\Xc_{\Sc}}.
\end{equation*} 
Hence, $d_{\min}(\Xc_{\pmb{a}_S}) \geq d_{\min}(\pmb{t}+\La_{\Xc_{\Sc}})=d_{\min}(\La_{\Xc_{\Sc}})$.

If a shortest vector of $\La_{\Xc_{\Sc}}$ lies in $M\Zb^K$, then $d_{\min}(\La_{\Xc_{\Sc}})=d_{\min}(M\Zb^K)=M$, and hence $d_{\min}(\Xc_{\pmb{a}_S}) \geq M$. This proves the second part of the lemma.
 
To prove the first part we will now show that \mbox{$d_{\min}(\Xc_{\pmb{a}_S}) \leq d_{\min}(\La_{\Xc_{\Sc}})$} if $\pmb{w}$ is a shortest vector of $\La_{\Xc_{\Sc}}$ and \mbox{$\pmb{w} \notin M\Zb^K$}. 
Note that \mbox{$\pmb{w} \mod M \neq \pmb{0}$} and \mbox{$\pmb{w} \mod M \in \Xc_{\Sc}$}. Hence, $d_{\min}\left(\Xc_{\Sc}\right) \leq \|\pmb{w} \mod M\| \leq \|\pmb{w}\|$. Since $\Xc_{\pmb{a}_S}$ is a coset of $\Xc_{\Sc}$ in $\Zb_M^K$, we have $d_{\min}(\Xc_{\pmb{a}_S})=d_{\min}(\Xc_{\Sc})$. Thus, we have
$d_{\min}(\Xc_{\pmb{a}_S}) = d_{\min}(\Xc_{\Sc}) \leq \|\pmb{w}\| = d_{\min}(\La_{\Xc_{\Sc}})$.
This completes the proof.
\end{IEEEproof}

Lemma~\ref{lem:dmin_lattice} provides the exact value of $d_S$, and hence 
$\sfrac{10\log_{10}\left(\sfrac{d_S^2}{d_0^2}\right)}{R_S}$, 
only if we can find a shortest vector \mbox{$\pmb{w} \in \La_{\Xc_{\Sc}}$} such that $\pmb{w} \mod M \neq \pmb{0}$. Otherwise, the lemma yields only a lower bound on $\sfrac{10\log_{10}\left(\sfrac{d_S^2}{d_0^2}\right)}{R_S}$.

\subsection{Computer search}

\begin{table}
\renewcommand{\arraystretch}{1.35}
\centering
\caption{Best Linear Index Codes with Circulant Encoding Matrix $\pmb{C}$.}
{\fontsize{6}{7}\selectfont{
\begin{tabular} {||c||c|c|c|c||}
\hline
\multirow{2}{*}{$M$} & \multicolumn{4}{c||}{$K=n$} \\
\cline{2-5}
                     & $2$ & $3$ & $4$ & $5$\\ 
\hhline{||=||====||}
 \multirow{2}{*}{$4$} & $(1,-2)$ & $(1,-2,-2)$ & $(1,1,-1,0)$ & $(1,-2,1,-1,0)$\\
   & $6.02$ & $4.52$ & $3.01$ & $3.76$ \\
\hline
 \multirow{2}{*}{$8$} & $(1,2)$ & $(1,2,0)$ & $(1,0,3,3)$ & $(1,-1,2,2,-3)$ \\
   & $4.65$ & $3.49$ & $4.01$ & $4.70$ \\
\hline
 \multirow{2}{*}{$16$}  & $(1,-4)$ & $(1,2,-6)$ & $(1,4,-6,-8)$ & $(1,-2,-5,-4,5)$\\
  & $6.02$ & $5.24$ & $5.57$ & $5.28$ \\
\hline
 \multirow{2}{*}{$32$}  & $(1,6)$ & $(1,-10,14)$ & $(1,10,14,2)$ & $(1,-8,-5,15,-6)$\\
  & $5.85$ & $5.73$ & $5.80$ & $5.77$ \\
\hline
 \multirow{2}{*}{$64$}  & $(1,-28)$ & $(1,-26,-4)$ & $(1,-26,20,30)$ & $(1,16,18,-9,21)$ \\
  & $6.04$ & $5.73$ & $5.85$ & $5.82$ \\
\hline
\end{tabular}
}}
\label{tbl:index_codes}
\end{table}

We use a computer search to find linear index codes with large side information gains. To reduce the complexity of the exhaustive search we restrict our search space to codes whose encoding matrices $\pmb{C}$ are circulant. We present results for \mbox{$n=K=2,3,4,5$} and \mbox{$M=4,8,16,32,64$}.
For each choice of $\pmb{C}$, with \mbox{$\det(\pmb{C}) \in U(\Zb_M)$}, we found that the value of $S$ that minimizes $\sfrac{10\log_{10}\left(\sfrac{d_S^2}{d_0^2}\right)}{R_S}$ yields a lattice $\La_{\Xc_{\Sc}}$ with a shortest vector $\pmb{w}$ such that $\pmb{w} \mod M \neq \pmb{0}$. Hence, using Lemma~\ref{lem:dmin_lattice}, we were able to calculate the exact value of \mbox{$\Gamma = \min_{S} \sfrac{10\log_{10}\left(\sfrac{d_S^2}{d_0^2}\right)}{R_S}$} for each candidate index code.
For each $M,K$, Table~\ref{tbl:index_codes} lists one index code with the largest side information gain $\Gamma$ among all codes with circulant encoding matrices. The table shows the first row of the circulant matrix $\pmb{C}$ and the side information gain $\Gamma$ (in~dB/b/dim).
All the index codes have \mbox{$\Gamma \geq 3$~dB/b/dim}, and for \mbox{$M \geq 16$}, the gain is at least $5.24$~dB/b/dim. 
In comparison, the codes from~\cite{NHV_arxiv_14} provide \mbox{$\Gamma \approx 6$~dB/b/dim}. Since the construction of~\cite{NHV_arxiv_14} relies on the Chinese remainder theorem, the resulting message sizes $|\Wc_1|,\dots,|\Wc_K|$ are powers of different primes. Here, we circumvent this problem by using codes over $\Zb_M$, but rely on numerical techniques to estimate $\Gamma$.

\section{Simulation Results \& Conclusion} \label{sec:4}

\begin{figure}[!t]
\centering
\includegraphics[totalheight=1.95in,width=3in]{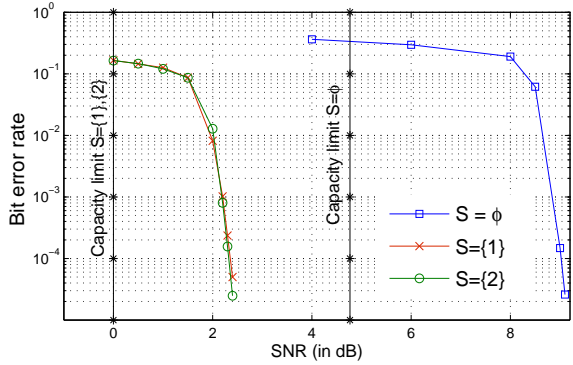}
\caption{Performance of the $16$-QAM index code used as a modulation scheme with two identical $(4000,2000)$ LDPC codes and iterative decoders.}
\label{fig:comparison_all}
\vspace{-5mm}
\end{figure} 

The proposed index codes are effective in exploiting receiver side information, but are sensitive to channel noise. The channel coding gain can be improved by encoding the $K$ information sources independently with channel codes, and modulating the resulting $K$ coded streams using a QAM index code. 
Consider \mbox{$K=2$} independent messages to be broadcast to three receivers, with $S=\varnothing,\{1\},\{2\}$, respectively. 
We use the $16$-QAM index code of Examples~\ref{ex:16QAM_1} and~\ref{ex:16QAM_2} (optimal from Table~\ref{tbl:index_codes}) concatenated with \mbox{$K=2$} identical \mbox{rate-$\sfrac{1}{2}$} $(4000,2000)$ regular LDPC codes (variable-node degree~3, check-node degree~6) catalogued in~\cite{Mac_Encyclopedia} using bit interleaved coded-modulation (BICM)~\cite{CTB_IT_98}. 
For each information source, $2000$ information bits are encoded into a $4000$ length LDPC codeword, which is then interleaved using a random interleaver. 
Four coded bits, two each from the two interleaved sequences, are mapped to two $\Zb_4$ symbols, which are then modulated to a $16$-QAM point using the index code of Example~\ref{ex:16QAM_1}.
The coded bit rate of each source is \mbox{$R_1=R_2=\sfrac{1}{2}$~b/dim}.
 
Each receiver regards the two information sources as independent users, and employs an iterative multiuser detector~\cite{Poo_SPMag_04} composed of three soft-in soft-out (SISO) a posteriori probability blocks~\cite{BDMP_CommLet_97}: one $16$-QAM demodulator, and two LDPC decoders. Each LDPC decoder block uses $50$ iterations between the check nodes and variable nodes, and the multiuser iterative demodulator-decoder uses $16$ iterations between the three SISO blocks.
For the receivers with $S=\{1\},\{2\}$, the side information is fed as input a~priori probabilities to the corresponding LDPC decoder.

From~\cite{Tun_IEEE_IT_06}, we know that a rate tuple $(R_1,R_2)$ is achievable if and only if \mbox{$\sfrac{1}{2}\log_2\left(1+{\sf SNR}\right) > \sum_{k=1}^{K}R_k - R_S$} for every receiver $({\sf SNR},S)$. For the three receivers corresponding to {$S=\varnothing,\{1\},\{2\}$}, $R_S$ equals $0$~b/dim, \mbox{$R_1=\sfrac{1}{2}$}~b/dim and \mbox{$R_2=\sfrac{1}{2}$~b/dim}, respectively. It follows that the minimum required ${\sf SNR}$ at the three receivers are $4.77$~dB, $0$~dB and $0$~dB, respectively.

Fig.~\ref{fig:comparison_all} shows the performance of the LDPC-coded $16$-QAM index code for \mbox{$S=\varnothing,\{1\},\{2\}$} and the capacity limits on the ${\sf SNR}$.
At bit error rate $10^{-4}$, the system performs $2.4$~dB from capacity for {$S=\{1\},\{2\}$}, and $4.3$~dB away for \mbox{$S=\varnothing$}.
While the LDPC code has contributed to channel coding gain, the symbol mapping provided by the inner index code has yielded significant ${\sf SNR}$ gains for the receivers that know either of the two messages a priori.


We have presented the first known family of index codes for the Gaussian broadcast channel that admit equal message rates, and with message sizes that are powers of $2$. The method employed to obtain these codes is limited to small values of $M$ and $K$ because of the complexity involved in the computer search. 
An analytical approach could extend the results to larger number of messages. Our simulations used a standard LDPC code designed for the single-user AWGN channel to improve noise resilience. Designing efficient coded-modulation techniques matched to the proposed modulation schemes may be crucial to achieve higher coding gains.


\end{document}